\documentstyle[a4,12pt,axodraw]{article}

\newcommand{\sm}{standard model}
\newcommand{\lsp}{lightest supersymmetric particle}
\newcommand{\xs}{cross section}
\newcommand{\br}{branching ratio}
\newcommand{\EW}{electroweak}
\newcommand{\vev}{vacuum expectation value}
\newcommand{\nwa}{narrow width approximation}
\newcommand{\susic}{supersymmetric}
\newcommand{\susy}{supersymmetry}
\newcommand{\sel}{selectron}
\newcommand{\trm}{transverse momentum}
\newcommand{\Sel}{\mbox{$\tilde e$}}
\newcommand{\snu}{sneutrino}
\newcommand{\Snu}{\mbox{$\tilde\nu$}}
\newcommand{\co}{chargino}
\newcommand{\no}{neutralino}
\newcommand{\No}{\mbox{$\tilde\chi^0_1$}}
\newcommand{\mpT}{\mbox{$p_{\perp}\!\!\!\!\!\!/\,\,\,$}}
\newcommand{\GeV}{\mbox{ GeV}}
\newcommand{\GEV}{{\rm GeV}}
\newcommand{\hra}{\hspace{1mm}\hookrightarrow}
\def\ep{$e^+e^-$}
\def\lo{leading order}

\def\PB{{\rm pb}}

\newcommand{\ee}{\mbox{$e^-e^-$}}
\newcommand{\eempt}{\mbox{$e^-e^-+\mpT$}}
\newcommand{\mmmpt}{\mbox{$\mu^-\mu^-+\mpT$}}
\newcommand{\eetoss}{\mbox{$e^-e^-\to\tilde e^-\tilde e^-$}}
\newcommand{\eetocc}{\mbox{$e^-e^-\to\tilde\chi_1^-\tilde\chi_1^-$}}

\def\gsim{\buildrel{\lower.7ex\hbox{$>$}}\over{\lower.7ex\hbox{$\sim$}}}
\def\lsim{\buildrel{\lower.7ex\hbox{$<$}}\over{\lower.7ex\hbox{$\sim$}}}

\begin{document}

\thispagestyle{empty}
\setcounter{page}{0}

\begin{flushright}
CERN-TH.6807/93\\
MPI-Ph/93-28\\
PSI-93-11\\
May 1993
\end{flushright}
\vspace*{\fill}
\begin{center}
{\Large\bf Supersymmetric Signals in $e^-e^-$ Collisions}$^*$\\
\vspace{2em}
\large
\begin{tabular}[t]{c}
Frank Cuypers$^{a,1}$\\
Geert Jan van Oldenborgh$^{b,2}$\\
Reinhold R\"uckl$^{a,c,d,3}$\\
\\
{$^a$ \it Sektion Physik der Universit\"at M\"unchen,
D--8000 M\"unchen 2, FRG}\\
{$^b$ \it Paul Scherrer Institut, CH-5232 Villigen PSI,
Switzerland}\\
{$^c$ \it Max-Planck-Institut f\"ur Physik,
Werner-Heisenberg-Institut,}\\
{\it D--8000 M\"unchen 40, FRG}\\
{$^d$ \it CERN,
CH-1211 Gen\`eve 23, Switzerland}\\
\end{tabular}
\end{center}
\vspace*{\fill}

\begin{abstract}
We consider the production and decay of \sel s and \co s
in \ee\ collisions.
The advantage over usual \ep\ collisions
is the very low level of \sm\ backgrounds
which should make the discovery of \sel s or \co s
relatively straightforward.
The use of polarized beams
provides an additional powerful tool
to determine the \susy\ parameters.
\end{abstract}

\vspace*{\fill}

\begin{flushleft}
\noindent$^1$ {\small Email: {\tt frank@hep.physik.uni-muenchen.de}}\\
\noindent$^2$ {\small Email: {\tt gj@csun.psi.ch}}\\
\noindent$^3$ {\small Email: {\tt rer@dmumpiwh.bitnet}}\\[12pt]
\noindent$^*$ {\footnotesize Work partially supported by the German Federal
Ministry for Research and Technology
under the contract No. 05 6MU93P
and by the CED Science Project No. SCI-CT 91-0729.}\\
\end{flushleft}

\newpage

\section{Introduction}

A large international effort is currently under way
to study the technical feasibility
and physics possibilities
of linear \ep\ colliders
in the TeV range.
A number of designs have already been proposed
(NLC, JLC, TESLA, CLIC, VLEPP,\dots)
and several workshops have recently been devoted to the subject.

{}From the experimental point of view
such a machine would provide exciting new possibilities,
including the use of highly polarized beams \cite{Damerell}
or the production of high energy photon beams \cite{LC12}.
Another
very useful option feasible
at a linear collider facility
consists in colliding electrons with electrons.
Of course,
in the realm of the \sm\
this option is not particularly interesting
because mainly M\o ller scattering and Bremsstrahlung events
are to be observed.
However,
it is just for that reason
that \ee\ collisions can provide unique information
on exotic processes, in particular on processes
involving lepton and/or fermion number violation.
For example,
heavy dileptons would yield an unmistakable signal \cite{F1},
Majorana neutrinos would strongly enhance the production of $W$ pairs
\cite{H1},
and \sel s may be abundantly produced \cite{LC11} and detected
through their decays.

We reexamine here this last possibility
and extend the analysis of Ref.~\cite{LC11}
to a full study of the \susy\ parameter space
with more realistic assumptions,
taking into account the knowledge accumulated in the past years.
In addition,
we also consider the production of \co s.
For definiteness,
we focus our analysis on a 500 GeV collider,
and indicate how the \xs s evolve with the energy.

In sections 2 and 3 we describe how \sel\ and \co\ pairs
can be produced in \ee\ collisions,
and give the relevant cross section formulae.
Section 4
is devoted to a short discussion of the decay modes of the \sel\ and
of \co s.
We then examine in section 5 the signals
and main \sm\ backgrounds
in unpolarized \ee\ collisions.
Finally, in section 6,
we show how the backgrounds can be reduced
to a very low level with the help of
polarized beams
and how information
on the values of the \susy\ parameters
can be gathered
that is complementary to what can be learned
{}from unpolarized tests.

\section{Selectron Production}

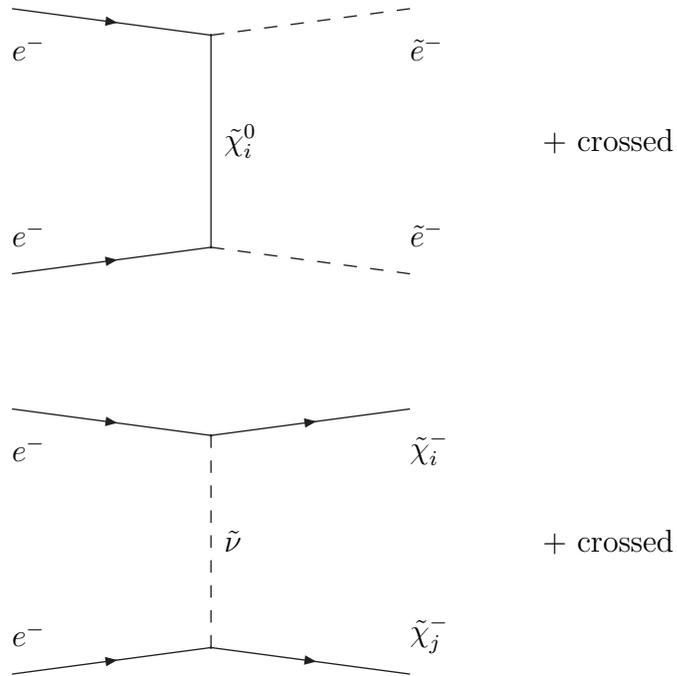
\begin{figure}[htb]
\begin{center}
\begin{picture}(250,150)(0,0)
\ArrowLine(0,0)(75,10)
\ArrowLine(0,100)(75,90)
\Vertex(75,90){.5}
\Line(75,90)(75,10)
\Vertex(75,10){.5}
\DashLine(75,10)(150,0){5}
\DashLine(75,90)(150,100){5}
\Text( 00,15)[l]{$e^-$}
\Text( 00,85)[l]{$e^-$}
\Text( 80,50)[l]{$\tilde\chi^0_i$}
\Text(150,85)[l]{$\tilde{e}^-$}
\Text(150,15)[l]{$\tilde{e}^-$}
\Text(200,50)[l]{+ crossed}
\end{picture}
\\
\begin{picture}(250,150)(0,0)
\ArrowLine(00,00)(75,10)
\ArrowLine(0,100)(75,90)
\Vertex(75,90){.5}
\DashLine(75,90)(75,10){5}
\Vertex(75,10){.5}
\ArrowLine(75,10)(150,0)
\ArrowLine(75,90)(150,100)
\Text( 00,15)[l]{$e^-$}
\Text( 00,85)[l]{$e^-$}
\Text( 80,50)[l]{$\tilde\nu$}
\Text(150,85)[l]{$\tilde{\chi}_i^-$}
\Text(150,15)[l]{$\tilde{\chi}_j^-$}
\Text(200,50)[l]{+ crossed}
\end{picture}
\end{center}
\caption[dummy]{Lowest order Feynman diagrams contributing to
\sel\ and \co\ production.}
\label{selfeyn}
\end{figure}

Selectron pairs can be produced in \ee\ collisions
by the u- and t-channel exchange of \no s,
as is shown on Fig.~\ref{selfeyn}.
Note that this reaction violates fermion number conservation,
which comes as no surprise since the \no s are Majorana fermions.

The yield of \sel s depends very crucially
on the properties of the exchanged \no s \cite{LC18},
{\em i.e.} their masses and their couplings to electrons,
because strong interferences can take place between the different channels
and dramatically influence the
production \xs. In the minimal model,
the masses $m_{\tilde\chi^0_i}$  of the four \no\ states (i=1,2,3,4)
and
their couplings $g_{iL,R}$ to electron-\sel\ pairs
of left(L) and right(R) chiralities
are complicated functions of four \susy\ parameters \cite{LC3}:
\begin{itemize}
\item   the ratio
$\tan\beta=v_2/v_1$
of the Higgs \vev s;
\item   the Higgs/higgsino superpotential mass parameter $\mu$;
\item   the soft \susy\ breaking mass parameter $M_1$
of the $U(1)_Y$ gaugino sector;
\item   the soft \susy\ breaking mass parameter $M_2$
of the $SU(2)_L$ gaugino sector.
\end{itemize}
In principle,
the mass parameters can take complex values.
Moreover,
the masses $m_{\tilde{e}_{L,R}}$
of the left- and right-\sel s
are arbitrary and need not be the same.
We are thus dealing with an appreciable number of degrees of freedom,
which can however be reduced by a few reasonable and
customary assumptions:
\begin{itemize}
\item   $\mu$ and $M_2$ are taken to be real.
\item   $M_1=5/3M_2\tan^2\theta_w$,
where $\theta_w$ is the weak mixing angle.
This is a consequence of the renormalization group evolution
{}from a common value $M_1=M_2$ at the GUT scale.
\item   All sleptons have the same mass
and are much lighter than the strongly interacting squarks and gluinos
$m_{\tilde\ell_L}=m_{\tilde\ell_R}=
m_{\tilde\nu_\ell}\ll m_{\tilde q},m_{\tilde g}$.
If renormalization group relations for scalar masses are used, this
corresponds to assuming a relatively large value for the common
scalar mass parameter $m_0$.
\item   The \lsp\ is a \no.
\item   R-parity remains unbroken.
Therefore,
\susic\ partners are always produced in pairs
and the lightest \no\ is stable
(in virtue of the previous assumption).
\end{itemize}
As it turns out,
for $\tan\beta\ge2$,
the dependence on $\tan\beta$ is generally weak.
For definiteness,
we thus set $\tan\beta=10$
in the following.
In contrast,
the results are very sensitive to the values of $\mu$ and $M_2$.
To cover the whole parameter space,
it is sufficient to consider positive and negative
values of $\mu$ and only positive values of $M_2$.
We shall later display our results in this $(\mu,M_2)$ half-plane.
For the time being,
however,
and for definiteness
we choose to work with $\mu=-300$ GeV and $M_2=300$ GeV.
In this case the masses of the lightest \no\ and \co\ are
$m_{\tilde\chi^0_1}=147$ GeV
and $m_{\tilde\chi^\pm_1}=255$ GeV.
These values lie well beyond the existing experimental constraints
\cite{LC24}.

The polarized differential \xs s for \sel\ production are given by
\begin{eqnarray}
{d\sigma(e_L^-e_R^-\to \tilde e_L^-\tilde e_R^-)\over dt}
&=
{\displaystyle{\pi\alpha^2\over s^2}
\sum_{i,j=1}^4}
&
g_{iL}^{\phantom*}~g_{iR}^{\phantom*}~g_{jL}^*~g_{jR}^*\
{tu-m_{\tilde e}^4
\over
\Bigl(m_{\tilde\chi^0_i}^2-t\Bigr)
\Bigl(m_{\tilde\chi^0_j}^2-t\Bigr)}
\ ,\label{dlr2ss}
\\
{d\sigma(e_L^-e_L^-\to \tilde e_L^-\tilde e_L^-)\over dt}
&=
{\displaystyle{\pi\alpha^2\over2s^2}
\sum_{i,j=1}^4}
&
{g_{iL}^{\phantom*}}^2{g_{jL}^*}^2\
s\, m_{\tilde\chi^0_i}\, m_{\tilde\chi^0_j}
\label{dll2ss}
\\
&&
\times
\left(
{1\over m_{\tilde\chi^0_i}^2-t}
+
{1\over m_{\tilde\chi^0_i}^2-u}
\right)
\left(
{1\over m_{\tilde\chi^0_j}^2-t}
+
{1\over m_{\tilde\chi^0_j}^2-u}
\right)
\ ,\nonumber
\end{eqnarray}
$\alpha\approx1/128$ being the fine structure constant, and
$s,t,u$ being the usual Mandelstam variables.
The formulas for the RL and RR \xs s
are obtained from the expressions above in an obvious manner.
The corresponding total \xs s are
\begin{eqnarray}
\sigma(e_L^-e_R^-\to \tilde e_L^-\tilde e_R^-)
&={\displaystyle{\pi\alpha^2\over s}\sum_{i,j=1}^4}&
g_{iL}^{\phantom*}~g_{iR}^{\phantom*}~g_{jL}^*~g_{jR}^*\
\biggl\{
-{\lambda\over s} - {1\over(m_{\tilde\chi^0_j}^2
-m_{\tilde\chi^0_i}^2)s}
\label{lr2ss}
\\
&&\hspace{-1cm}
\times
\Bigl[
\bigl(m_{\tilde\chi^0_i}^2s+(m_{\tilde\chi^0_i}^2-m_{\tilde
 e}^2)^2\bigr)L_i
-\bigl(m_{\tilde\chi^0_j}^2s+(m_{\tilde\chi^0_j}^2-m_{\tilde
 e}^2)^2\bigr)L_j
\Bigr]
\biggr\}
\ ,\nonumber
\end{eqnarray}
\begin{eqnarray}
\sigma(e_L^-e_L^-\to \tilde e_L^-\tilde e_L^-)
&={\displaystyle{\pi\alpha^2\over s}\sum_{i,j=1}^4}
&{g_{iL}^{\phantom*}}^2{g_{jL}^*}^2\
{m_{\tilde\chi^0_i} m_{\tilde\chi^0_j}
\over
(m_{\tilde\chi^0_j}^2-m_{\tilde\chi^0_i}^2)
(s+m_{\tilde\chi^0_i}^2+m_{\tilde\chi^0_j}^2-2m_{\tilde e}^2)}
\label{ll2ss}
\\
&&\quad
\times
\Bigl[(s+2m_{\tilde\chi^0_j}^2-2m_{\tilde
 e}^2)L_i-(s+2m_{\tilde\chi^0_i}^2-2m_{\tilde e}^2)L_j\Bigr]
\ ,\nonumber
\end{eqnarray}
where
\begin{equation}
\lambda=\lambda(s,m_{\tilde e}^2,m_{\tilde e}^2)
=\sqrt{s^2-4m_{\tilde e}^2s}
\label{lambda}
\end{equation}
and
\begin{equation}
L_i=\ln{s+2m_{\tilde\chi^0_i}^2-2m_{\tilde e}^2+\lambda\over
 s+2m_{\tilde\chi^0_i}^2-2m_{\tilde e}^2-\lambda}
\ .\label{log}
\end{equation}
For $i=j$ the limit of these expressions has to be taken carefully
and agrees with the results obtained for the exchange
of a single photino in Ref.~\cite{LC11}.

\begin{figure}[t]
\centerline{
\begin{picture}(554,504)(0,0)
\put(0,0){\strut\epsffile{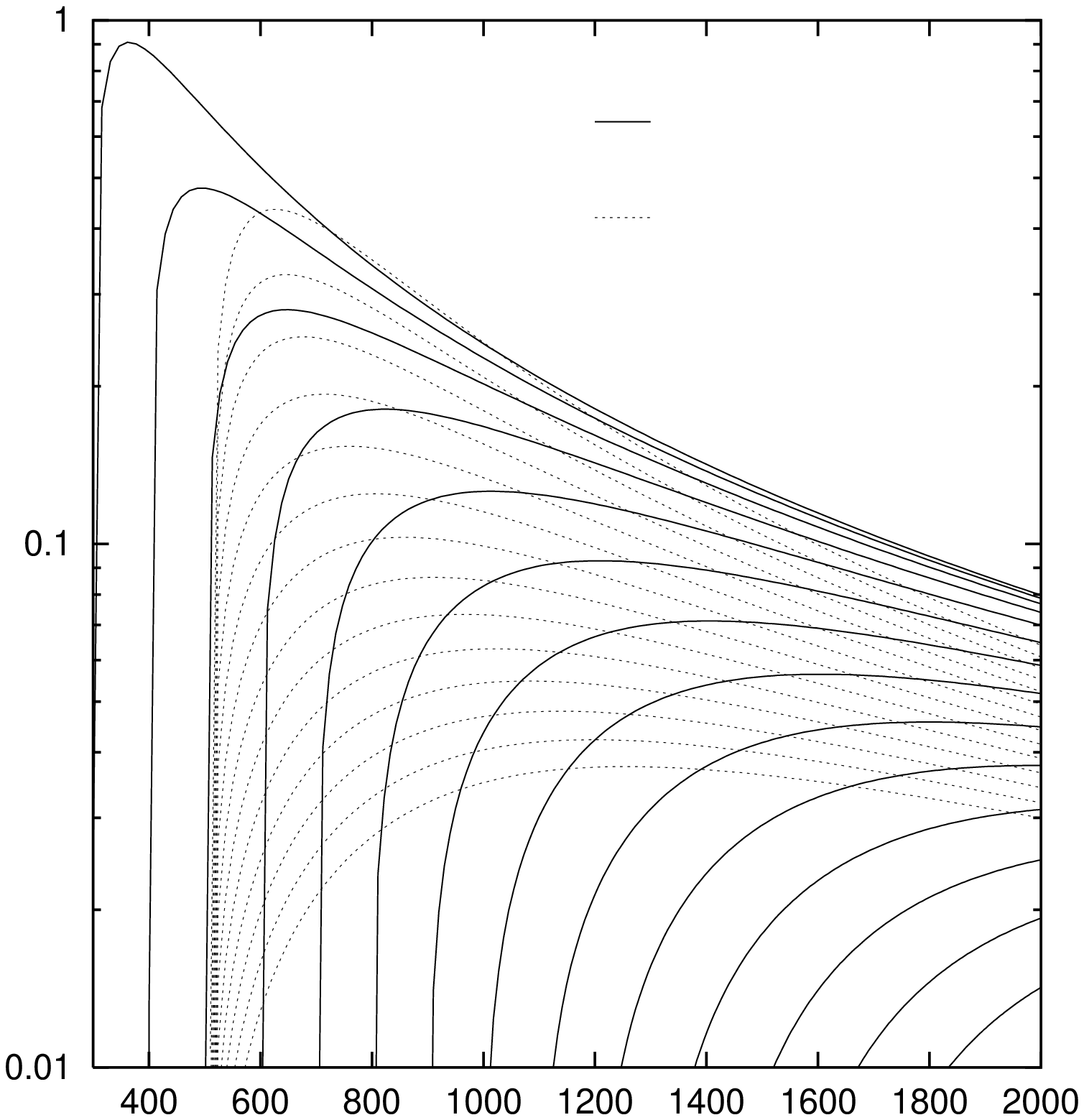}}
\put( 70.8,468.9){\makebox(0,0)[tr]{\Large$\sigma[\PB]$}}
\put(480.8, 19.1){\makebox(0,0)[tr]{\Large$\sqrt{s}[\GEV]$}}
\Text(350,430)[l]{\large$e^-e^-\to\tilde e^-\tilde e^-$}
\Text(350,390)[l]{\large$e^-e^-\to\tilde\chi_1^-\tilde\chi_1^-$}
\end{picture}
}
\caption[dummy]{Energy dependence of the unpolarized production \xs s of
$e^-e^-\to\tilde e^-\tilde e^-$ (full curves)
and
$e^-e^-\to\tilde\chi_1^-\tilde\chi_1^-$ (dotted curves)
for $m_{\tilde e}=m_{\tilde \nu}=$150, 200, \ldots\ 800 GeV,
assuming
$\tan\beta=10$, $\mu=-300$ GeV and $M_2=300$ GeV.
For this choice of parameters,
$m_{\tilde\chi^-_1}=255$ GeV.
}
\label{eny}
\end{figure}

\begin{figure}[t]
\centerline{
\begin{picture}(554,504)(0,0)
\put(0,0){\strut\epsffile{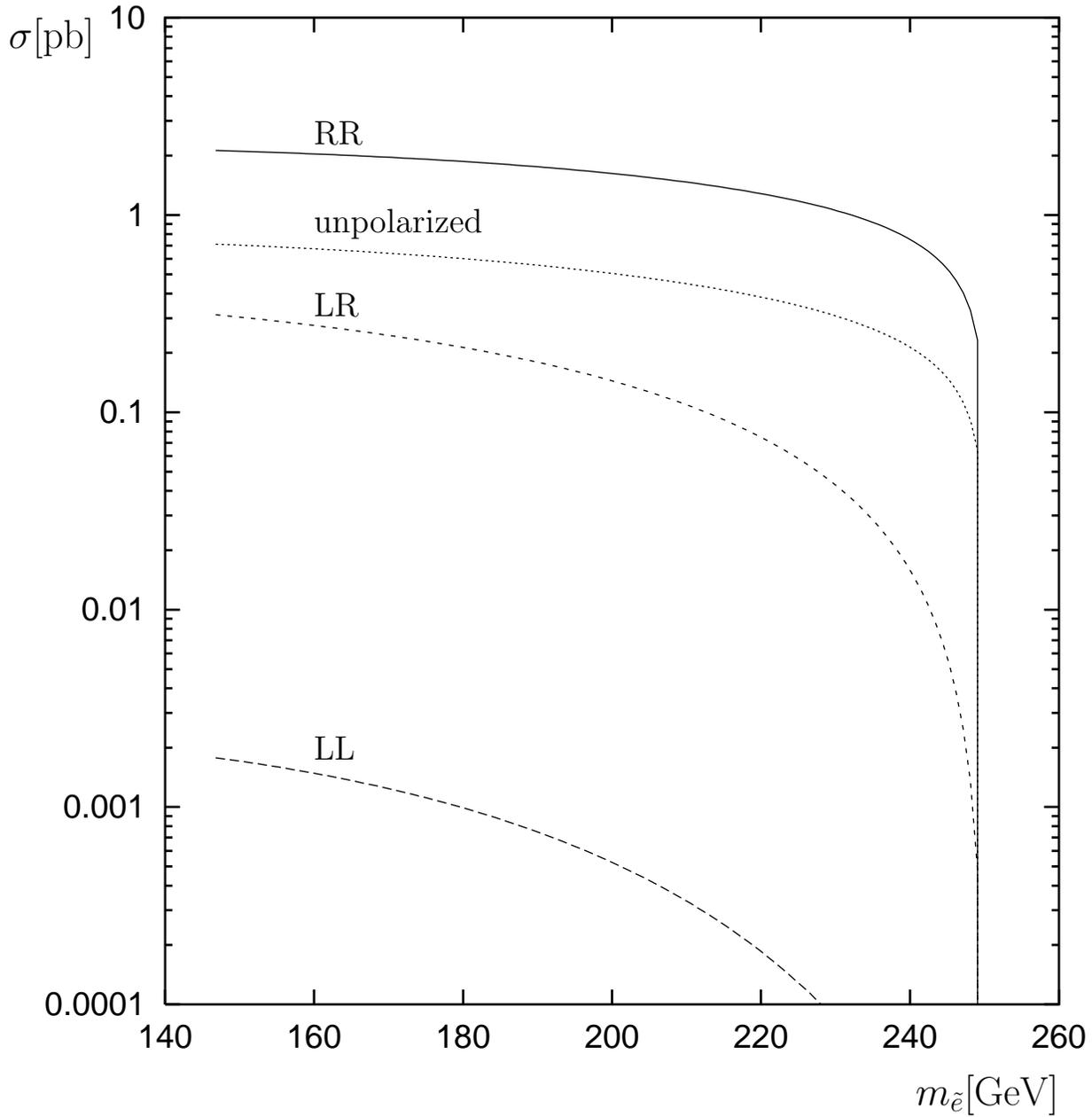}}
\put( 70.8,468.9){\makebox(0,0)[tr]{\Large$\sigma[\PB]$}}
\put(480.8, 19.1){\makebox(0,0)[tr]{\Large$m_{\tilde e}[\GEV]$}}
\Text(164.1,419.8)[l]{\large RR}
\Text(164.1,381.1)[l]{\large unpolarized}
\Text(164.1,346.7)[l]{\large LR}
\Text(164.1,158.3)[l]{\large LL}
\end{picture}
}
\caption[dummy]{Dependence of the total production \xs s
for \sel\ pairs
on the mass of the \sel\
in polarized and unpolarized $e^-e^-$ scattering at $\sqrt{s}=500$ GeV.
The choice of \susy\ parameters is the same as in Fig.~\ref{eny},
in particular, $m_{\tilde e_L}=m_{\tilde e_R}=m_{\tilde e}$.}
\label{mass}
\end{figure}

The energy dependence of the \sel\ production \xs\
is shown in Fig.~\ref{eny}
for several \sel\ masses.
Clearly,
the highest \xs s are obtained just above threshold.
Away from threshold the dependence on the \sel\ mass is quite weak,
as can be seen more clearly in Fig.~\ref{mass}
where we have plotted the dependence of the \sel\ production \xs\
as a function of the \sel\ mass
for unpolarized as well as for polarized electron beams
and for the centre-of-mass energy $\sqrt{s}=500$ GeV.

\section{Chargino Production}

Chargino pairs can be produced in \ee\ collisions
by the u- and t-channel exchange of a \snu,
as is also shown in Fig.~\ref{selfeyn}.
Since \co s only couple to left-handed leptons,
only the LL component of a given $e^-e^-$ initial state contributes.
The polarized differential \xs s are given by
\begin{eqnarray}
{d\sigma(e_L^-e_L^-\to \tilde\chi_i^-\tilde\chi_j^-)\over dt}
&=&
{\displaystyle{1\over1+\delta_{ij}}\
{\pi\alpha^2\over s^2}}
|g_ig_j|^2
\\
&&
\times
\Biggl\{
        {\Bigl(m_{\tilde\chi^-_i}^2-t\Bigr)
         \Bigl(m_{\tilde\chi^-_j}^2-t\Bigr)
        \over
         \bigl(m_{\tilde\nu}^2-t\bigr)^2}
        +
        {\Bigl(m_{\tilde\chi^-_i}^2-u\Bigr)
         \Bigl(m_{\tilde\chi^-_j}^2-u\Bigr)
        \over
         \bigl(m_{\tilde\nu}^2-u\bigr)^2}
\nonumber
\\
&&
\hspace{-3cm}
        -\
        {\Bigl[
                \Bigl(m_{\tilde\chi^-_i}^2-t\Bigr)
                \Bigl(m_{\tilde\chi^-_j}^2-t\Bigr)
                +
                \Bigl(m_{\tilde\chi^-_i}^2-u\Bigr)
                \Bigl(m_{\tilde\chi^-_j}^2-u\Bigr)
                -
                s\Bigl(s-m_{\tilde\chi^-_i}^2-m_{\tilde\chi^-_j}^2\Bigr)
         \Bigr]
        \over
         \bigl(m_{\tilde\nu}^2-t\bigr)\bigl(m_{\tilde\nu}^2-u\bigr)}
\Biggr\}
\ .\label{dll2cc}
\nonumber
\end{eqnarray}
The masses $m_{\tilde\chi^-_i}$ of the \co s
and
their couplings to \sel s $g_i$
are also functions of the three \susy\ parameters
$\tan\beta$, $\mu$ and $M_2$.
For the total \xs\ one obtains
\begin{eqnarray}
\sigma(e_L^-e_L^-\to \tilde\chi_i^-\tilde\chi_j^-)
&=&
{\displaystyle{1\over1+\delta_{ij}}\
{2\pi\alpha^2\over s^2}}
|g_ig_j|^2
\label{ll2cc}
\\
&&
\hspace{-4cm}
\times
\Biggl\{
        ~2\lambda
        ~+~
        \left[
                \Pi
                -
                m_{\tilde\nu}^2\Sigma
                +
                m_{\tilde\nu}^4
        \right]
        ~F~
        ~+~
        {\Sigma^2+2\Pi+6m_{\tilde\nu}^4-6m_{\tilde\nu}^2\Sigma
        +4m_{\tilde\nu}^2s-s\Sigma
        \over
         \Sigma-2m_{\tilde\nu}^2-s}
        ~L~
\Biggl\}
\ ,\nonumber
\end{eqnarray}
where
\begin{eqnarray}
\Sigma
&=&
m_{\tilde\chi_i^-}^2+m_{\tilde\chi_j^-}^2
\\
\Pi
&=&
m_{\tilde\chi_i^-}^2m_{\tilde\chi_j^-}^2
\\
\lambda
=
\lambda(s,m_{\tilde\chi^-_i}^2,m_{\tilde\chi^-_j}^2)
&=&
\sqrt{   s^2
        +m_{\tilde\chi^-_i}^4
        +m_{\tilde\chi^-_j}^4
        -2sm_{\tilde\chi^-_i}^2
        -2sm_{\tilde\chi^-_j}^2
        -2m_{\tilde\chi^-_i}^2m_{\tilde\chi^-_j}^2}
\\
F
&=&
{\lambda\over m_{\tilde\nu}^4+m_{\tilde\nu}^2(s-\Sigma)+\Pi}
\\
L
&=&
\ln
{s+2m_{\tilde\nu}^2-\Sigma+\lambda
\over
s+2m_{\tilde\nu}^2-\Sigma-\lambda}
\ .
\end{eqnarray}

The dependence of the unpolarized \xs\
$\sigma(e^-e^-\to\tilde\chi_1^-\tilde\chi_1^-)$
on the collider energy
is shown in Fig.~\ref{eny}\ for several values of
the mass of the exchanged
 \snu.
We repeat that
this plot has been obtained for
$\tan\beta=10$, $\mu=-300$ GeV and $M_2=300$ GeV,
{\em i.e.} for a \co\ mass $m_{\tilde\chi_1^-}=255$ GeV.

\section{Decays of the Selectron and Charginos}

Since all the (s)particles considered here
decay through \EW\ interactions,
their lifetimes are typically long
in comparison to their mass scale.
It is therefore
safe to use the \nwa,
which we will do in the following.

The simplest decay mode of the \sel\
is into an electron and the lightest \no:
\begin{equation}
        \tilde e^-\to e^-\No~.
\label{seldec}
\end{equation}
Since we assume the \no\ to be the \lsp,
only the electron is visible.
In the \eetoss\ reaction,
this decay yields a clean \eempt\ event
which we suggest to use as a signal
for the production of \sel s.

If kinematically allowed,
other decays can take place in addition, most importantly,
\begin{eqnarray}
\tilde e^-
&\to&
e^-\tilde\chi^0_2~,
\label{seldec1}\\
&\to&
\nu_e\tilde\chi^-_1~,
\label{seldec2}
\end{eqnarray}
and similar decays into the heavier neutralino and chargino states.
The \susic\ particles produced in this way
will decay into lighter (s)particles,
which themselves might undergo further decays
until only conventional particles and a number of
the lightest \no\ remain.
The end-product of such cascade decays
can sometimes again be an electron
accompanied by invisible particles only.
For the \eetoss\ reaction,
this mechanism can thus
provide an important enhancement of the \eempt\ signal
in regions of the \susic\ parameter space
where the direct decay (\ref{seldec}) is not dominant \cite{LC18}.
To compute the \br\ for the decay
$\Sel\to e^-+$invisible,
we shall make use of the two-body decay algorithm
described in \cite{LC18}.

The decays of \co s are even more complicated.
Concentrating on the lightest \co,
if kinematically allowed to do so,
it will decay into leptons and sleptons or $W$'s and \no s:
\begin{eqnarray}
\tilde\chi^-_1
&\to&
\ell^-\Snu_\ell~,
\label{codec1}
\\
&\to&
\tilde\ell^-\nu_\ell~,
\label{codec2}
\\
&\to&
W^-\tilde\chi_i^0
\ .\label{codec3}
\end{eqnarray}
For simplicity,
we discard the possibility
\begin{eqnarray}
\tilde\chi^-_1
&\to&
H^-\tilde\chi_i^0
\label{codec4}
\end{eqnarray}
by assuming the charged Higgs boson to be too heavy.
If the \co\ is heavier than the sleptons,
it will preferentially decay
with approximately a 50\%\ \br\ in each of the channels
(\ref{codec1}) and (\ref{codec2})
and with democratic
probabilities for the different flavours \cite{LC17}.
In this case,
the sleptons can only decay further
into leptons and \no s
and will eventually yield a
$\ell^-\ell^-+\mpT$ signal
in the \eetocc\ reaction.
If the \co\ is lighter than the sleptons
but can still decay according to the reaction (\ref{codec3}),
one has to deal with a $W^-W^-$ signal.
It can also happen that none of the two-body decays
(\ref{codec1}-\ref{codec4})
is kinematically allowed.
If this is the case
and if the mass of the sleptons is much larger than the mass of the $W$,
the decay through a virtual $W$ dominates.
The \br\ of the leptonic decay
$\tilde\chi^-_1\to\ell^-\bar\nu_\ell\No$
is then approximately 42\%\
\cite{LC14}.
In the next section,
we will concentrate our analysis
of the \co\ production in \ee\ collisions
on the \mmmpt\ signal.

\section{Unpolarized Electron Beams}

\begin{figure}[htb]
\begin{center}
\begin{picture}(250,150)(0,0)
\ArrowLine(00,00)(75,00)
\ArrowLine(0,80)(75,80)
\Vertex(75,80){.5}
\Photon(75,80)(75,00){5}{5}
\Vertex(75,00){.5}
\ArrowLine(75,00)(150,0)
\ArrowLine(75,80)(150,80)
\Vertex(100,80){.5}
\Photon(100,80)(160,130){5}{5}
\Text( 00,10)[l]{$e^-$}
\Text( 00,70)[l]{$e^-$}
\Text( 85,40)[l]{$\gamma,Z^0$}
\Text(150,70)[l]{$e^-$}
\Text(150,10)[l]{$e^-$}
\Text(170,120)[l]{$Z^0$}
\Text(200,50)[l]{$+\ \cdots$}
\end{picture}
\\
\begin{picture}(250,150)(0,0)
\ArrowLine(00,00)(75,00)
\ArrowLine(0,80)(75,80)
\Vertex(75,80){.5}
\Photon(75,80)(75,00){5}{5}
\Vertex(75,00){.5}
\ArrowLine(75,00)(150,0)
\ArrowLine(75,80)(150,80)
\Vertex(100,80){.5}
\Photon(100,80)(160,130){5}{5}
\Text( 00,10)[l]{$e^-$}
\Text( 00,70)[l]{$e^-$}
\Text( 85,40)[l]{$\gamma,Z^0$}
\Text(150,70)[l]{$\nu_e$}
\Text(150,10)[l]{$e^-$}
\Text(170,120)[l]{$W^-$}
\Text(200,50)[l]{$+\ \cdots$}
\end{picture}
\\
\begin{picture}(250,200)(0,-50)
\ArrowLine(00,00)(75,00)
\ArrowLine(0,80)(75,80)
\Vertex(75,80){.5}
\Photon(75,80)(75,00){5}{5}
\Vertex(75,00){.5}
\ArrowLine(75,00)(150,0)
\ArrowLine(75,80)(150,80)
\Vertex(100,80){.5}
\Photon(100,80)(160,130){5}{5}
\Vertex(100,0){.5}
\Photon(100,0)(160,-30){5}{5}
\Text( 00,10)[l]{$e^-$}
\Text( 00,70)[l]{$e^-$}
\Text( 85,40)[l]{$\gamma,Z^0$}
\Text(150,70)[l]{$\nu_e$}
\Text(150,10)[l]{$\nu_e$}
\Text(170,120)[l]{$W^-$}
\Text(170,-20)[l]{$W^-$}
\Text(200,50)[l]{$+\ \cdots$}
\end{picture}
\end{center}
\caption[dummy]{Lowest order Feynman diagrams contributing to
the \sm\ background processes considered.}
\label{backfeyn}
\end{figure}
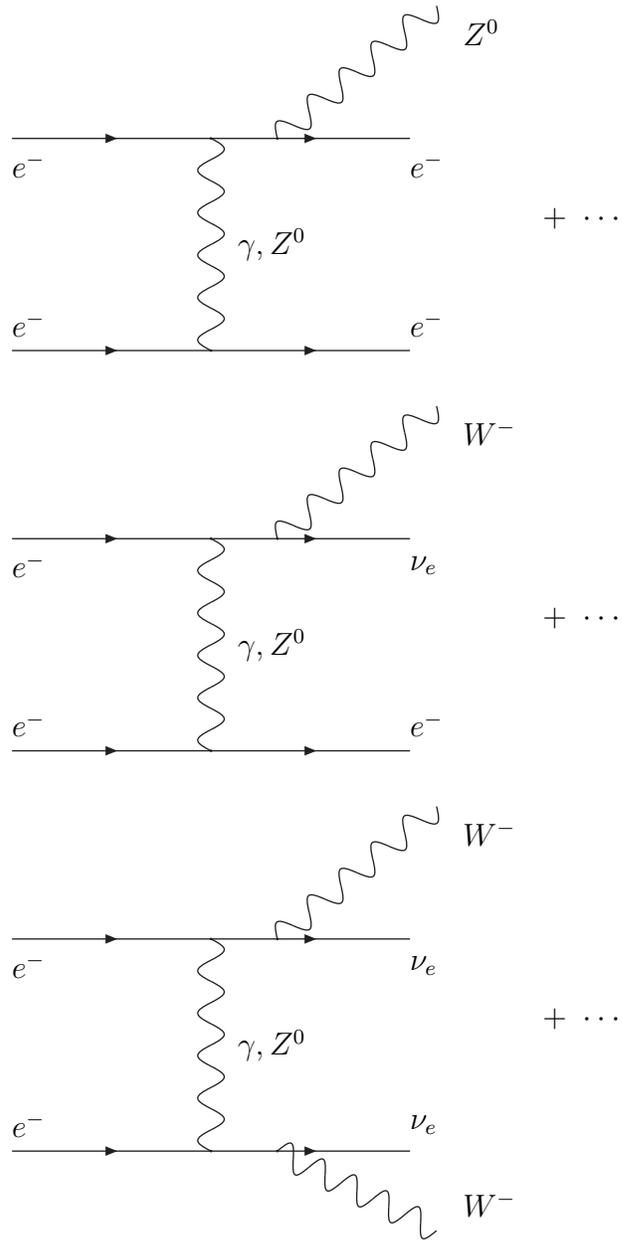

To select \eempt\ events containing the \sel\ signal
we impose the following kinematical cuts
on the two observed electrons:
\begin{itemize}
\item   the rapidity cut
        \begin{equation}
                |\eta_e| < 3~;
        \label{rapcut}
        \end{equation}
\item   the energy cut
        \begin{equation}
                E_e > 5 \GeV~;
        \label{enycut}
        \end{equation}
\item   the acoplanarity cut
        \begin{equation}
                ||\phi(e^-_1)-\phi(e^-_2)|-180^\circ| > 2^\circ~,
        \label{phicut}
        \end{equation}
        where $\phi$ is the azimuthal angle of the electrons
        with respect to the beam axis.
\end{itemize}
The two cuts (\ref{rapcut},\ref{enycut})
are supposed to guarantee good detector acceptance.
The acoplanarity cut (\ref{phicut}) is designed to eliminate
the abundant M\o ller electron pairs.
The leading \sm\ backgrounds
which then remain
originate from $W^-$ and $Z^0$ Bremsstrahlung:
{\arraycolsep0cm
\renewcommand{\arraystretch}{0}
\begin{equation}
        e^-e^- \to
        \begin{array}[t]{ll}
                e^-\nu_e&W^- \qquad ,\\
                      &\hra e^-\bar\nu_e
        \end{array}
\label{enw}
\end{equation}
\begin{equation}
        e^-e^- \to
        \begin{array}[t]{ll}
                e^-e^-&Z^0 \qquad .\\
                      &\hra\nu\bar\nu
        \end{array}
\label{eez}
\end{equation}
}
These backgrounds,
whose typical \lo\ Feynman diagrams are depicted in Fig.~\ref{backfeyn},
are quite sensitive to the cuts (\ref{rapcut}-\ref{phicut})
because their \xs s are dominated by large
collinear logarithms.
After cuts and including the relevant branching ratios,
the \xs s are 150 fb for $W^-$ Bremsstrahlung (\ref{enw})
and 40 fb for $Z^0$ Bremsstrahlung (\ref{eez}).

\begin{figure}[t]
\centerline{
\begin{picture}(554,504)(0,0)
\put(0,0){\strut\epsffile{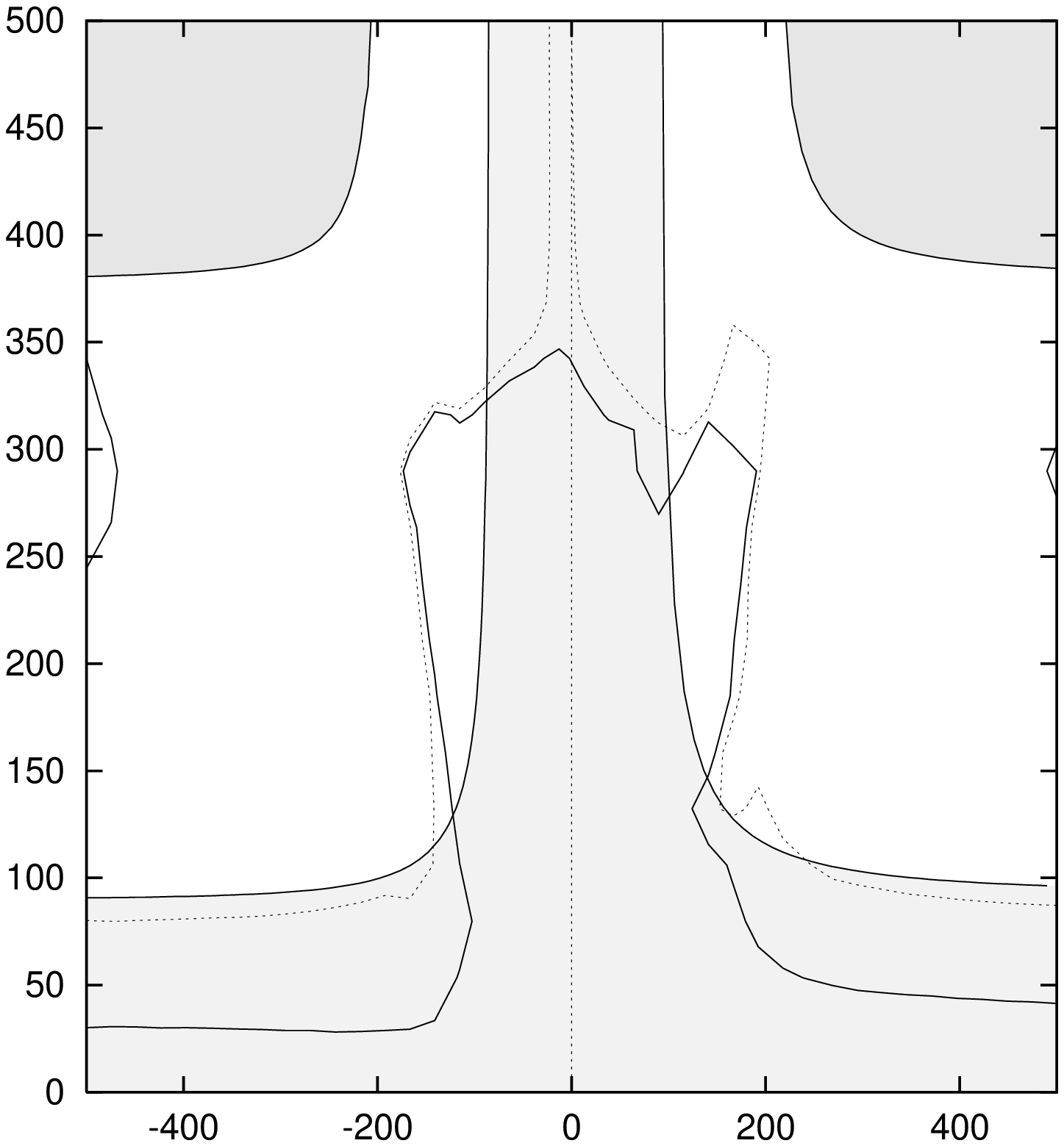}}
\Text(70,460)[tr]{\Large$M_2$[GeV]}
\Text(481,19)[tr]{\Large$\mu$[GeV]}
\Text(291,133)[b]{\Large LEP2}
\Text(112.2,426.9)[lb]{\Large unphysical}
\Text(389.6,426.9)[lb]{\Large unphysical}
\Text(116,293)[lb]{\Large 1 pb}
\Text(475,293)[rb]{\Large 1 pb}
\Text(363,217)[lb]{\Large 0.1 pb}
\end{picture}
}
\caption[dummy]{Contours in the \susy\ parameter space
corresponding to the unpolarized \xs s
$\sigma(e^-e^- \rightarrow \tilde e^- \tilde e^- \rightarrow \eempt)
=$ 0.1 and 1 pb for $m_{\tilde e}=$ 200 GeV, $\tan\beta=$10,
and the cuts defined in (\ref{rapcut}-\ref{phicut}).
The regions labelled `unphysical' are excluded since there
$m_{\tilde e} < m_{\tilde\chi^0_1}$ in contradiction to our
assumption of the $\chi^0_1$ being the lightest \susy\ particle.
The contours which would be obtained if cascade decays are ignored
are also shown with dotted lines.
}
\label{npss}
\end{figure}

On the other hand,
the acceptance cuts (\ref{rapcut},\ref{enycut})
have almost no impact on the \susic\ \eempt\ signal
{}from \sel\ production,
except for the very small region of parameter space where
$m_{\tilde e}-m_{\tilde\chi_1^0}<10$ GeV
and the decay electrons have little energy.
Also the acoplanarity cut (\ref{phicut})
reduces the signal by at most 3\%.
In Fig.~\ref{npss} we have displayed the contours
in the $(\mu,M_2)$ half-plane
where the signal \xs\
for a 200 GeV \sel\
reaches 0.1 and 1 pb,
respectively.
Comparing this result with the above background estimates
we conclude that
over a large part of the parameter space beyond the region
which will be explored by LEP2,
the signal to background ratio is of order
one or more.
Moreover,
the 1 pb contour cuts through an area of parameter space
where there are no cascade decays.
In contrast,
the 0.1 pb contour is located in a region
where cascade decays are important
and have to be taken into account.
However,
for the moderate energy cut (\ref{enycut})
most of those electrons which emerge at the end of a cascade
like the ones from
$\tilde {\chi}^0_2$ and $\tilde {\chi}^-_1$ decays
in (\ref{seldec1}) and (\ref{seldec2}),
still carry enough energy
to be observable.
We have also outlined
in Fig.~\ref{npss}
the contours which are obtained
when only the direct decay (\ref{seldec}) is considered
and cascade decays of the type (\ref{seldec1},\ref{seldec2})
are neglected.

Here we should point out that
in order to obtain the \br s for cascade decays,
we have used a two-body decay algorithm \cite{LC18}
which yields imprecise results
when some of the two-body decays become kinematically impossible
and three-body decay formulae \cite{LC22} have to be used.
This happens when
$m_{\chi^0_2}<\min(m_{\tilde\ell},m_{\chi^0_1}+m_Z)$
or $m_{\chi^-_1}<\min(m_{\tilde\ell},m_{\chi^0_1}+m_W)$
as is the case for low values of $\mu$ or $M_2$.
Our contours including cascade decays might thus not be too precise.
However, they certainly cannot reach beyond the dotted contours.
Since most of this region of parameter space
will already be explored by LEP2 anyway,
this uncertainty is not really relevant.

\begin{figure}[t]
\centerline{
\begin{picture}(554,504)(0,0)
\put(0,0){\strut\epsffile{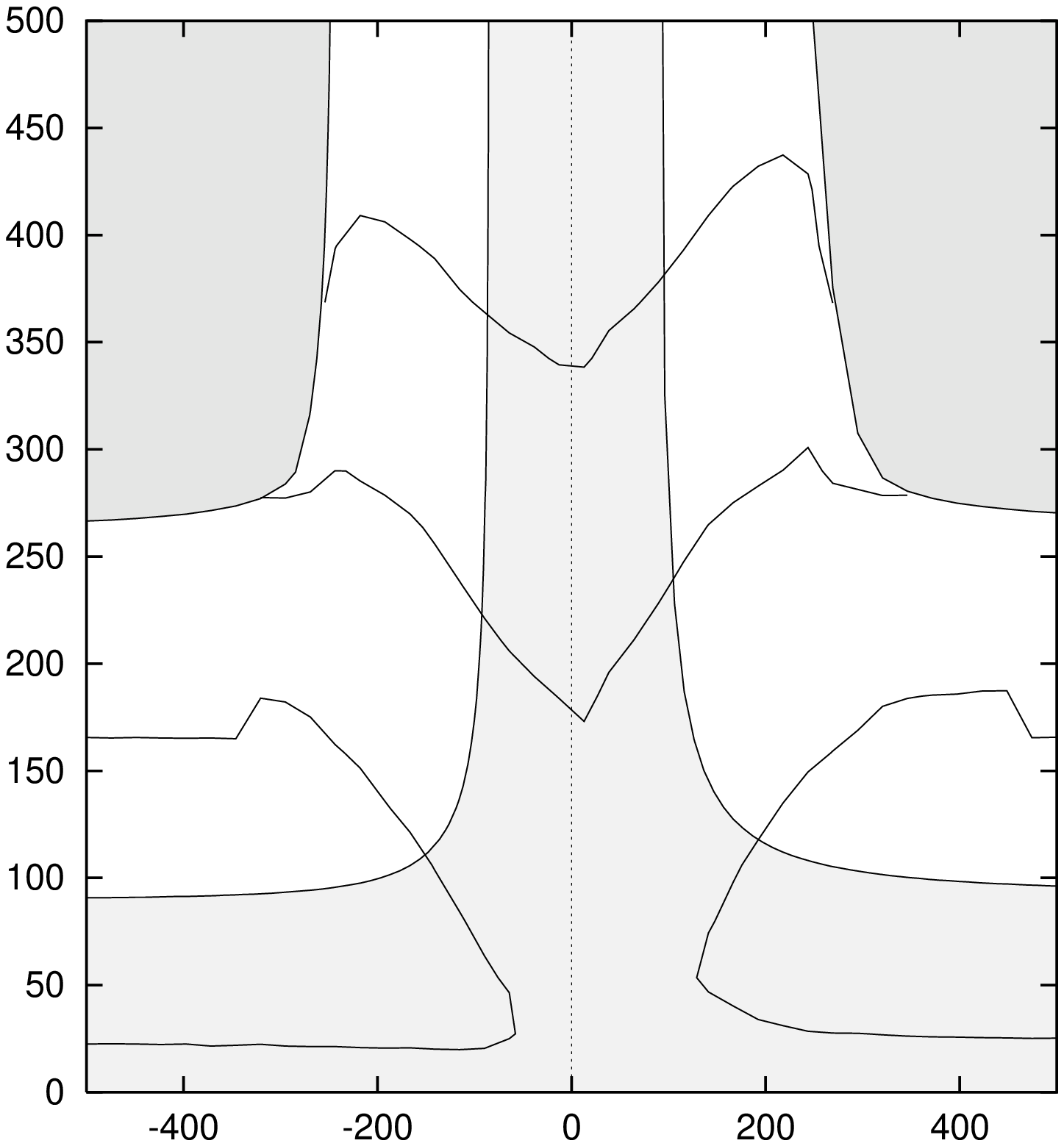}}
\Text(70,460)[tr]{\Large$M_2$[GeV]}
\Text(481,19)[tr]{\Large$\mu$[GeV]}
\Text(291,133)[b]{\Large LEP2}
\Text(177,217)[l]{\Large 100 fb}
\Text(196,309)[l]{\Large 10 fb}
\Text(215,406)[l]{\Large 1 fb}
\Text(148,406)[b]{\Large no pair-}
\Text(148,377)[b]{\Large production}
\Text(433,406)[b]{\Large no pair-}
\Text(433,377)[b]{\Large production}
\end{picture}
}
\caption[dummy]{Contours in the \susy\ parameter space
corresponding to the unpolarized \xs s
$\sigma(e^-e^- \rightarrow
\tilde \chi^-_1 \tilde \chi^-_1 \rightarrow \mmmpt)=$
1, 10 and 100 fb including
the cuts (\ref{rapcut}-\ref{phicut}).
The chargino mass $m_{\tilde \chi^-_1}$
varies with $M_2$ and $\mu$, while the sneutrino mass is set to
300 GeV and $\tan\beta=$10.}
\label{npcc}
\end{figure}

If \sel s are too heavy to be produced in pairs,
\co s can still save the day.
Concentrating on the \mmmpt\ signal,
the background arises mainly from double $W$ Bremsstrahlung (\ref{ww}):
{\arraycolsep0cm
\renewcommand{\arraystretch}{0}
\begin{equation}
        e^-e^- \to
        \begin{array}[t]{ll}
                W^-\nu_e&W^-\nu_e \qquad .\\
                        &\hra\mu^-\bar\nu_\mu\\
                \multicolumn{2}{l}{\hra\mu^-\bar\nu_\mu}
        \end{array}
\label{ww}
\end{equation}
}
This process is very laborsome to evaluate.
An order of magnitude estimate suggests that the \xs\ should be
less than 10\% of the \xs\ obtained for the
single $W$ Bremsstrahlung process (\ref{enw}),
{\em i.e.} approximately 100 fb.
Consequently,
the \xs\ for obtaining
a \mmmpt\ event from this \sm\ source within the cuts
should not exceed 1 fb.
Concentrating on \co\ production and
assuming 300 GeV for the
mass of the exchanged \snu\,
we have plotted in Fig.~\ref{npcc}
the contours in the $(\mu,M_2)$ half-plane
along which the observable \xs\ for the \mmmpt\ signal
{}from the decay modes (\ref{codec1},\ref{codec2})
is 1, 10 and 100 fb.
We find that as long as $M_2\lsim300$ GeV,
the signal to background (\ref{ww}) ratio should comfortably exceed one.

\section{Polarized Electron Beams}

In \co\ pair production,
polarization can only enhance the signal
(as well as the background)
by a factor up to four.
For \sel s,
however,
the use of polarized electron beams
completely changes the picture.

\begin{figure}[t]
\centerline{
\begin{picture}(554,504)(0,0)
\put(0,0){\strut\epsffile{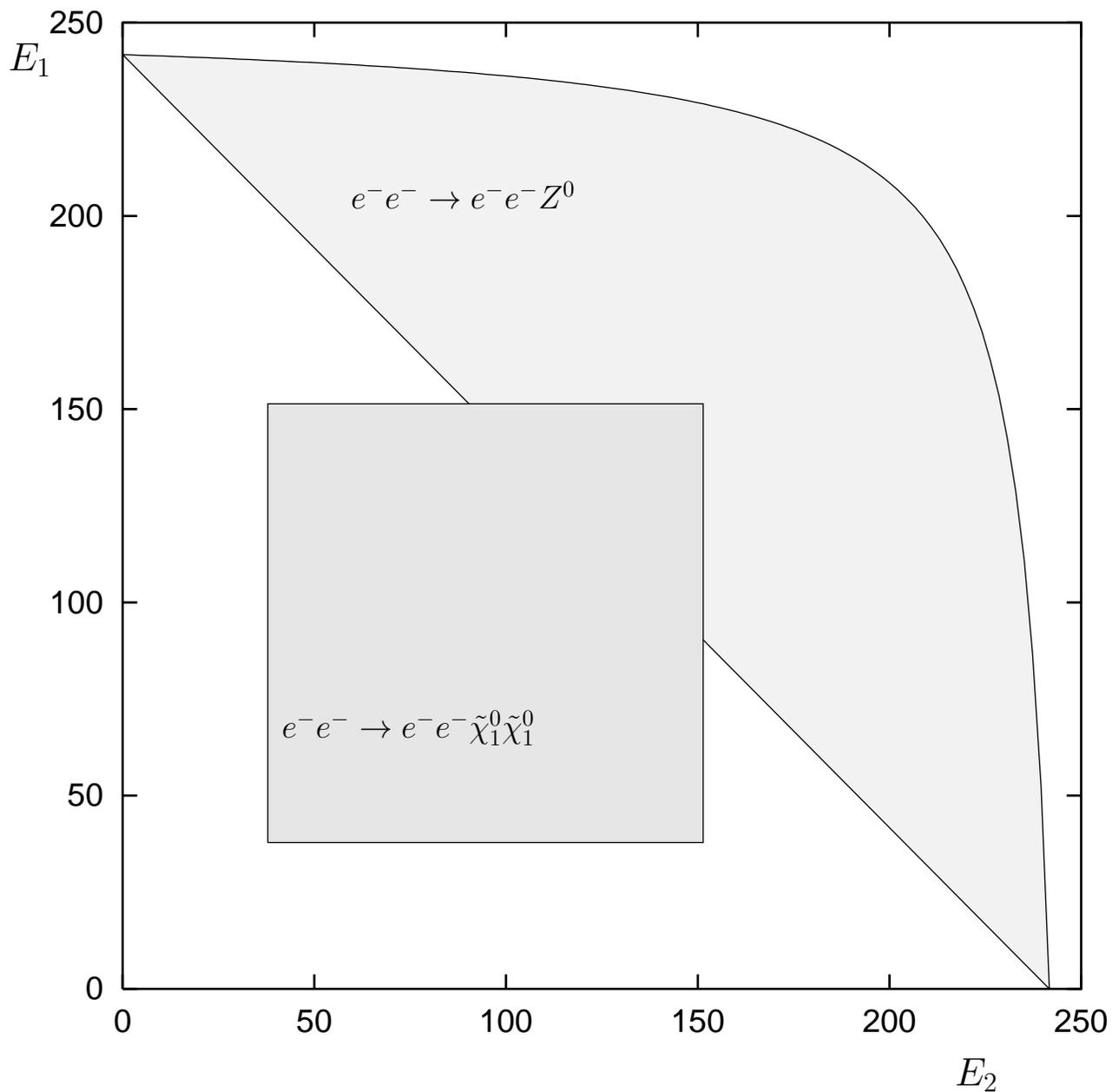}}
\Text(70,460)[tr]{\Large$E_1$}
\Text(481,19)[tr]{\Large$E_2$}
\Text(200,400)[tl]{\large$e^-e^-\to e^-e^-Z^0$}
\Text(170,170)[tl]{\large$e^-e^-\to e^-e^-\tilde\chi^0_1\tilde\chi^0_1$}
\end{picture}
}
\caption[dummy]{Allowed range of energies
for the final state electrons in the processes
$e^-e^-\to e^-e^-Z^0$
and
$e^-e^-\to\tilde e^-\tilde e^-\to e^-e^-\tilde\chi^0_1\tilde\chi^0_1$.
For the latter reaction
we have assumed $m_{\tilde e}=$ 200 GeV
and $m_{\tilde \chi^0_1}=$ 100 GeV.
}
\label{ps}
\end{figure}

Besides a possible enhancement of the \eempt\ signal
for some values of the \susy\ parameters,
working with right-handed
polarized beams has the immense advantage
of eliminating the most important background
{}from $W^-$ Bremsstrahlung (\ref{enw}).
In this situation,
it is then worthwhile to try to
also eliminate the background
{}from $Z^0$ Bremsstrahlung (\ref{eez}),
in order to select a really clean sample of \susic\ events
with no,
or negligibly little,
background from \sm\ processes.
This is indeed possible
since the energies $E_1$ and $E_2$ of the electron pairs
which originate from $Z^0$ Bremsstrahlung
and from \sel\ production and decay
are very distinctly distributed in the phase space.
The respective boundaries are given by
\begin{eqnarray}
\begin{array}{l}
        E_1+E_2 ~>~ {\displaystyle{s-m_Z^2\over2\sqrt{s}}} \\
        \strut\\
        (\sqrt{s}-2E_1)(\sqrt{s}-2E_2) ~>~ m_Z^2
\end{array}
\label{zps}
\end{eqnarray}
for $e^-e^-\to e^-e^-Z^0$, and by
\begin{eqnarray}
\begin{array}{l}
        E_1,E_2 ~>~ {\displaystyle{\sqrt{s}\over4}}
        \left(
                1-{m^2_{\tilde\chi^0_1}\over m^2_{\tilde e}}
        \right)
        \left(
                1-\sqrt{1-{4m^2_{\tilde e}\over s}}
        \right)\\
        \strut\\
        E_1,E_2 ~<~ {\displaystyle{\sqrt{s}\over4}}
        \left(
                1-{m^2_{\tilde\chi^0_1}\over m^2_{\tilde e}}
        \right)
        \left(
                1+\sqrt{1-{4m^2_{\tilde e}\over s}}
        \right)\
\end{array}
\label{sps}
\end{eqnarray}
for $e^-e^-\to\tilde e^-\tilde e^-\to e^-e^-\No\No$.
A typical example of such boundaries
is shown in the Dalitz plot of Fig.~\ref{ps}.
As it can be seen,
by imposing the high energy cut
\begin{equation}
E_{e_1}+E_{e_2} < {s-m_Z^2\over2\sqrt{s}} \approx 242 \GeV
\label{e1e2cut}
\end{equation}
one can eliminate all electron pairs
which originate from $Z^0$ Bremsstrahlung without affecting much
the signal.
At worst 55\%\ of the electron pairs
which originate from \sel\ production
can be lost by this cut,
and for many values of the \sel\ and \no\ masses
it has no incidence.

\begin{figure}[t]
\centerline{
\begin{picture}(554,504)(0,0)
\put(0,0){\strut\epsffile{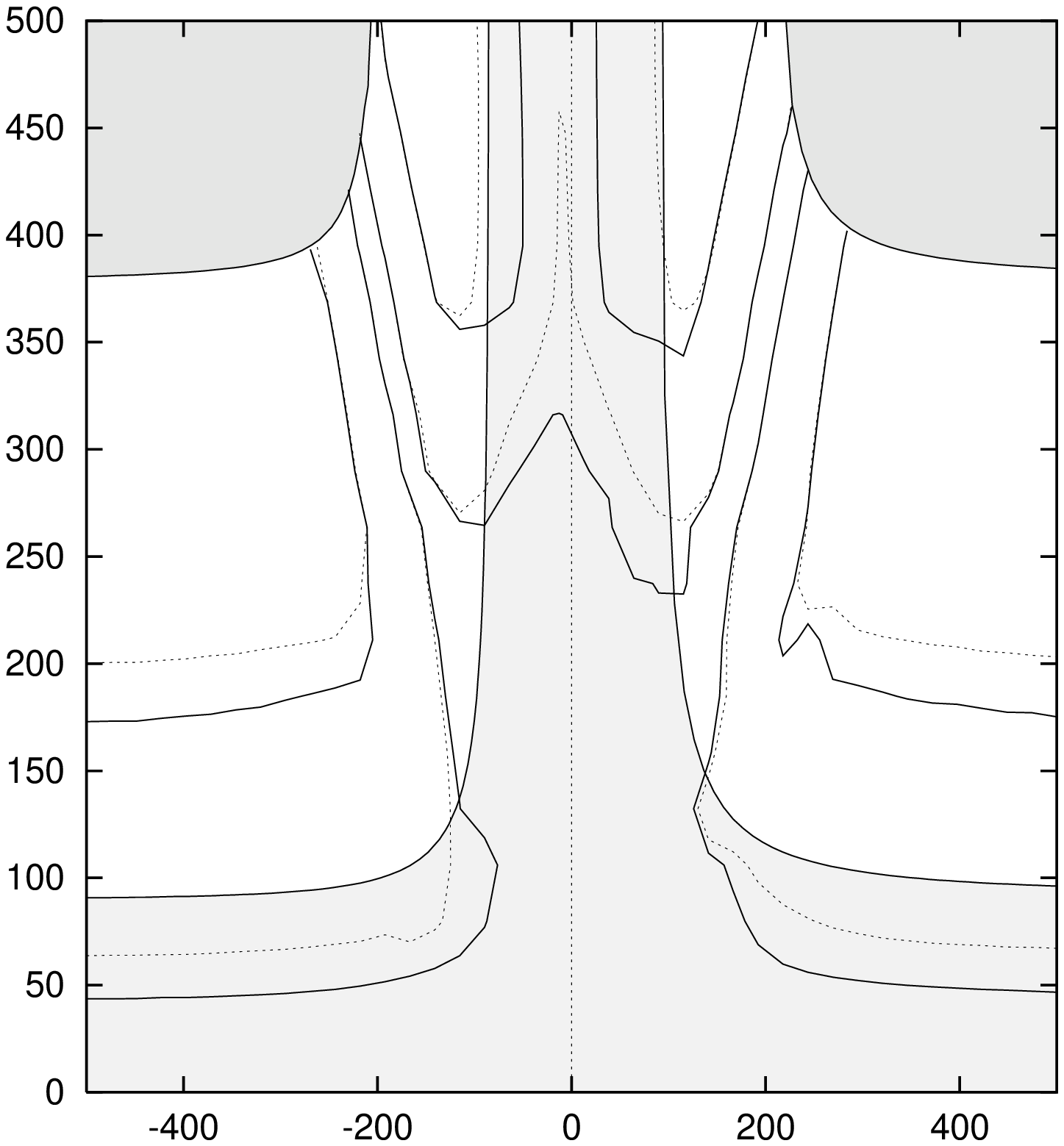}}
\Text(70,460)[tr]{\Large$M_2$[GeV]}
\Text(481,19)[tr]{\Large$\mu$[GeV]}
\Text(291,133)[b]{\Large LEP2}
\Text(112.2,426.9)[lb]{\Large unphysical}
\Text(389.6,426.9)[lb]{\Large unphysical}
\Text(207,259)[r]{\Large 1 pb}
\Text(386,259)[l]{\Large 1 pb}
\Text(222,452)[l]{\Large 1 pb}
\Text(359,452)[r]{\Large 1 pb}
\Text(239,172)[r]{\Large 0.1 pb}
\Text(352,182)[l]{\Large 0.1 pb}
\Text(291,325)[b]{\Large 0.1 pb}
\end{picture}
}
\caption[dummy]{Same as Fig.~\ref{npss},
for electron beams with right-handed polarization
and including the energy cut (\ref{e1e2cut}).
The contours which would be obtained if cascade decays are ignored
are indicated by dotted lines.
}
\label{rrss}
\end{figure}

The overall discovery potential of such a polarization experiment
is illustrated in Fig.~\ref{rrss}.
As in Fig.~\ref{npss} for the unpolarized case
we display the 0.1 pb and 1 pb contours in the $(\mu,M_2)$ half-plane.
The contours
which would be obtained
if cascade decays are neglected,
are outlined by the dotted curves.
Outside of the area which will be covered by LEP2
the influence of cascades is only marginal.
In comparison to the unpolarized case,
the area where the \xs\ exceeds 1 pb
has grown noticeably.
On the other side,
there are now two bands
with small \xs s (less than 0.1 pb)
along the lines $M_2=2|\mu|$
which are quite narrow but which
reach deeply into the unexplored $(\mu,M_2)$ regions.
This effect is caused
by subtle interferences between
the different \no s exchanged in the u- and t-channel.
A similar destructive interference is observed
along the lines $M_2=|\mu|$
in the case of left-handed polarization.
This also explains the huge differences in \xs\
for different polarizations
observed already in Fig.~\ref{mass}.
Obviously, this characteristic feature can be used
to determine or, at least, constrain
the \susy\ parameters
by comparing the \eempt\ yield
for different combinations of polarizations of the electron beams.

Furthermore,
if the upper and lower limits $E_{max}$ and $E_{min}$
of the electron energies in the signal events (see (\ref{sps}))
can be determined more or less accurately,
it should be possible to derive rough values for
both the mass of the \sel\ and the mass of the lightest \no:
\begin{equation}
\begin{array}{lcl}
        m_{\tilde e}
        &=&
        \sqrt{s}
        {\displaystyle\sqrt{E_{max}E_{min}\over(E_{max}+E_{min})^2}}\\
        \strut\\
        m_{\tilde\chi^0_1}
        &=&
        m_{\tilde e}
        {\displaystyle\sqrt{1-{2(E_{max}+E_{min})\over\sqrt{s}}}}\ .
\end{array}
\label{masses}
\end{equation}
Note that this mass determination
is of purely kinematical nature.
It does not depend on the values taken by any of the remaining
\susy\ parameters
and is thus entirely model-independent.
The effects of smearing due to
initial state Bremsstrahlung should be further investigated.

\section{Conclusions}

Because of the low level of \sm\ backgrounds,
\ee\ collisions are an ideal reaction for discovering
and investigating \susy\ at linear colliders.
The \sel\ and \co\ production \xs s
are of the same order as in \ep\ collisions
but the backgrounds are down by more than one order of magnitude.
Moreover,
in contrast to \ep\ \cite{LC23}, $e^-\gamma$ \cite{LC16}
or $\gamma\gamma$ \cite{LC17} collisions,
possible cascade decays of the \sel\ would be clearly observed,
since no strong \trm\ cuts are necessary here
to differentiate a \susic\ \eempt\ signal from the \sm\ background.

For \sel\ production,
right-handed polarization of the
electron beams can further enhance the signal to background ratio
and in principle allow for a direct,
model-independent
measurement of the \sel\ and \no\ masses.
In addition,
comparison of \xs s
for different polarizations
can provide information
on the values taken by the various
gaugino/higgsino mixing parameters.

\bigskip
\bigskip
\bigskip
\noindent
Part of the \xs s we used have been computed with the help
of the dedicated computer algebra program CompHEP \cite{CHEP1}.
We are very much indebted to Edward Boos and Mishael Dubinin
for having provided us with this software.

%\bibliography{/user/frank/Papeles/bibliography}
%\bibliographystyle{unsrt}

\newpage

\section{Figures}

\vspace{-2cm}

\vspace{-2cm}

\end{document}